\documentclass[onecolumn,amsmath,amssymb,aps]{revtex4}

\usepackage{graphicx}
\usepackage{dcolumn}
\usepackage{bm}
\usepackage[utf8x]{inputenc}
\usepackage{float}

\begin{document}


\title{$X(3872)$ and Its Heavy Quark Spin Symmetry Partners in QCD Sum Rules}

\author{Halil Mutuk}
 \email{halilmutuk@gmail.com}
 \affiliation{Physics Department, Ondokuz Mayis University, Samsun, Turkey}

\author{Yasemin Saraç}
\affiliation{Electrical and Electronics Engineering Department, Atilim University, Ankara, Turkey}

\author{Hasan Gümüş}
 \affiliation{Physics Department, Ondokuz Mayis University, Samsun, Turkey
}

\author{Altuğ Özpineci}
\affiliation{Physics Department, Middle East Technical University, Ankara, Turkey}

\begin{abstract}
$X(3872)$ presents many surprises after its discovery more than ten years ago. Understanding its properties is crucial to understand  the spectrum of possible exotic mesons. In this work,  $X(3872)$ meson and its heavy quark spin symmetry (HQSS) partners (including the mesons in the bottom sector) are studied within the QCD Sum Rules approach using a current motivated by the molecular picture of $X(3872)$. We predict four heavy partners to $X(3872)$ and bottomonium with the masses and $J^{PC}$ quantum numbers. Obtained results are in good agreement with the previous studies and available experimental data within errors. 
\end{abstract}

\pacs{Later}
\maketitle

\section{\label{sec:level1}Introduction}
During the last decade experimental data from Belle, BaBar, BESIII, CDF, D0, LHCb and other collaborations have put some puzzles about the conventional hadron spectrum, $i.e.$, quark-antiquark and three-quarks picture of hadrons. New states (called exotic states) were observed from these experiments and they opened a new era of hadron spectroscopy. 

The milestone of these exotic states is $X(3872)$. The $X(3872)$ was first observed by Belle Collaboration in 2003 and the production mode was $B^+ \rightarrow X(3872)K^+\rightarrow J/\psi \pi^+ \pi^-K^+$ \cite{1}. It has been also confirmed by the CDF \cite{2}, D0 \cite{3}, and BaBar \cite{4} collaborations. The current average mass of $X(3872)$ is $3871.69 \pm 0.17$ MeV and it is only $0.16$ MeV below  the $D^0\bar{D}^{*0}$ threshold with a less then $1.2$ MeV full width \cite{5}. Its quantum numbers were determined by the LHCb Collaboration to be $J^{PC}=1^{++}$ in 2013 \cite{6}. Its unusual properties presents a puzzle in the meson spectroscopy and up to now there is no consensus about its structure. 

The natural attempt to investigate $X(3872)$ is using quark model as a $c\bar{c}$ state. According to quark model, $X(3872)$ is a $2P$ charmonium state. The mass of this state was obtained as 3947 MeV in \citep{7,8} and 3906 MeV in \cite{9}. In \cite{10} they studied $X(3872)$ resonance as $c\bar{c}=\chi_{c_1}(2P)$ which sits on the $D^{*0}\bar{D}^0$ threshold and has a mass of $m(D^{*0}\bar{D}^0)=3871.81 \pm 0.36 $ MeV. They also suggest a program for experimental research in order to verify their assumption. The other quark model candidates with  $J^{PC}=1^{++}$ are 2$^3P_1(3925)$ and 3$^3P_1(3853)$ \citep{11,12}. Lattice QCD calculations give  2$^3P_1(4010)$ \cite{13} and  2$^3P_1(4067)$ \cite{14}. As can be seen, the masses of 2$^3P_1$ charmonium state are bigger than the observed state.

The other inconsistency with quark model is the  $J/\psi \pi^+ \pi^-$ and $J/\psi \pi^+ \pi^- \pi^0$ decays. The combined result for ratio of the decay fractions of $X(3872)$ into $J/\psi \pi^+ \pi^-$ and $J/\psi \pi^+ \pi^- \pi^0$ is \cite{15}
\begin{equation}
\frac{Br( X(3872) \to J/\psi \pi^+ \pi^- \pi^0)}{Br(X(3872)\to J/\psi \pi^+ \pi^-)}= 0.8 \pm 0.3. 
\end{equation}
In these decays, the pions are produced through the decay of intermediate $\rho$ or $\omega$ mesons, respectively. If one considers the differences in phase space between $\rho$ and $\omega$ mesons, the production amplitude ratio can be found as \cite{16}
\begin{equation}
\vert \frac{A(J/\psi \rho)}{A(J/\psi \omega)}   \vert = 0.26 \pm 0.007.
\end{equation}
Such a big isospin violation cannot be occur in the quark model since from the study of dipion mass distribution in the $X(3872) \to J/\psi \pi^+ \pi^- $ decay, Belle \cite{1} and CDF \cite{17} concluded that decay process proceeds through $X(3872) \to J\psi \rho$. Due to a charmonium state has isospin zero, it cannot decay easily into a $ J\psi \rho$ final state \cite{18}. However, such a isospin violation is possible in the molecular picture, due to the mass difference of $D^{(*)0}$ ve $D^{(*)\pm}$ \cite{19}.

Another interesting observation about $X(3872)$ is its  radiative decays. The branching ratio of $X(3872)$ to $J/\psi $ and $\psi (2S)$ is \citep{20,21}
\begin{equation}
\frac{Br(X\rightarrow \psi (2S)\gamma )}{Br(X\rightarrow J/\psi \gamma )}%
=2.46\pm 0.64\pm 0.29.
\end{equation}
It is a question why $X(3872)$ prefers to decay into $\psi(2S) \gamma$ even though the phase space is much smaller than its decay into $J/\psi \gamma$. In the charmonium picture, $X\rightarrow \psi (2S)\gamma$ is a $\Delta L=1$ transition.  It is claimed in \cite{22} that this ratio cannot be explained naturally in a pure molecular picture. However it is stated that this ratio can be obtained by adding a charmonium admixture into molecular picture \cite{23,24,25}. 

The properties of $X(3872)$ made it difficult to reconcile with a pure $c\bar{c}$ state in quark model picture \cite{26}. Thus several methods have been proposed to study properties of $X(3872)$ which include a radial excitation of the $P$-wave charmonium \cite{27}, a tetraquark \cite{28}, a mixture of an ordinary charmonium and a hadronic molecule \citep{26,29}, a state generated in the coupled-channel dynamical scheme \cite{30,31}, heavy quark effective field theory approach \cite{32,33,34,35}.  In \cite{36}, the authors made a threshold parametrization of the Belle and BaBar data on $B$ decays to $KJ/\psi \pi^+ \pi^-$ and $KD\bar{D}^{*0}$. They showed that data can be reproduced with a similar quality for $X(3872)$ being a bound and/or virtual state. They also noted that $X(3872)$ can be a higher order virtual state pole in the limit in which the small $D^{*0}$ width vanishes. Among these models, a popular description of $ X(3872)$ is as a molecular state consisting $D$ and $\bar{D}^*$ \cite{37,38,39,40}.

A natural framework to study  hadronic states with heavy quarks is presented by the Heavy Quark Spin Symmetry (HQSS). In the limit where the masses of heavy quarks are taken to infinity, or in other words 
$\Lambda_{QCD}/m_Q$ where $m_Q$ denotes quark mass, the spin of the quark decouples from the dynamics which refers the strong interactions in the system are independent of the heavy quark spin. This implies that the states that differ only in the spin of the heavy quark, $i.e.$ states in which the rest of the system has the same total angular momentum, should be degenerate. By using HQSS, it is obtained that if $X(3872)$ is a bound state of $D$ and $\bar{D}^*$, then it should have degenerate partners in heavy quark limit \citep{33,34}. The same result was achieved in \cite{41} by means of the heavy quark limit of QCD. They concluded that using interpolating currents which they proposed, the states that couple to them form degenerate triplets with the quantum numbers $J^{PC}=2^{++}$, $J^{PC}=1^{++}$, and $J^{PC}=0^{++}$. They also reported that in heavy quark limit this conclusion holds for any state that couples to the currents independent of its internal structure. In \citep{42,43}, the authors studied $X(3872)$ by using HQSS with the assumption of hadronic molecule. 

In the molecular picture of $X(3872)$, the two charm quarks have a total spin $S_H=1$ (also $S_H=0$ can be in molecules), and the light quarks have a total spin $S_l=1$. The total spin of such a system, assuming $L=0$, can be $J=0$, $J=1$ and $J=2$. In the heavy quark limit, all these three states should be degenerate. In addition, the state in which $S_H=0$ and $S_l=1$ should also be degenerate with the previous three, forming a heavy quark spin symmetry quartuplet having the quantum numbers $J^{PC}=0^{++}$, $1^{++}$, $2^{++}$ and $1^{-+}$.  Note also that, in the heavy quark limit for both the $c$ and $b$ quarks, there appears a flavor symmetry between these two quarks.  Using this symmetry, it is possible to extract information about the $c$ sector using the $b$ sector, and vice versa.

In the framework of QCD Sum Rules (QCDSR), the mass and current-meson coupling constant of the exotic $X(3872)$ state are computed within the two-point sum rule method using the diquark-antidiquark and molecule interpolating currents \cite{44}. In \cite{45}, they studied $X(3872)$ meson with thermal QCDSR method. Aliev $et ~ al.$ investigated $X(3872)$ via QCD sum rules as a mixing of charmonium and molecular $D^*D$ states and found that $Y(3940)$, $X(4260)$ and their orthogonal combinations should exist \cite{46}. Azizi and Er investigated $X(3872)$ in cold nuclear matter using diquark-antidiquark current within the framework of the in-medium two-point QCD sum rule method. They found that the mass, current-meson coupling and vector self energy strongly depend on the density of cold nuclear matter \cite{47}. Other QCDSR studies on $X(3872)$ can be found in \cite{48,49,50,51,52,53}. In \cite{54}, they studied color singlet-singlet type and octet-octet type currents to interpolate the $X(3872)$, $Z_c(3900)$ and $Z_b(10610)$ and concluded that more theoretical and experimental works are still needed to distinguish the molecule and tetraquark assignments; while there are no candidates for the color octet-octet type molecular states.

In \cite{41}, currents to be used to study $X(3872)$ and its partners in a QCD sum rules framework has been proposed. In this work, $X(3872)$ and its $J^{PC}=0^{++}$ and $J^{PC}=2^{++}$ partners are studied within a QCDSR framework using the currents proposed in \cite{41}. Obtained sum rules are also used to study the corresponding mesons in the bottom sector.

The paper is organized as follows. In section \ref{sec:level2}, QCDSR is given briefly and the degeneracy of the $X(3872)$ is obtained by QCD Sum Rules method. Section \ref{sec:level3} is devoted to numerical results of this degeneracy and in section \ref{sec:level4} we summarize our results. 

\section{\label{sec:level2}QCD Sum Rules and Degeneracies From the Correlation Functions}

QCD Sum Rules are formulated by Shifman, Vainsthein and Zakharov in 1979 \cite{55} for mesons and generalized to baryons by Ioffe \cite{56} in 1981. It is one of the celebrated method among non-perturbative methods such as lattice QCD, AdS/QCD, Chiral Perturbation Theory etc. The method is based on the study of a suitable chosen correlation function in two different kinematical regions.

On one side, it is calculated in the deep Euclidean region where the correlation function receives dominant contribution from short distances. In this case, the correlation function can be calculated using operator product expansion (OPE). On the other side, one calculates the correlation function for positive momentum squared. In this kinematical region, the correlation function can be expressed in terms of the properties of the hadrons. The two expression are matched using spectral representation of the
correlation function, and hadronic properties are extracted by this matching.

The fundamental object of the QCD Sum Rule is the correlation function
\begin{equation}
\Pi(q^2)=i \int d^4x e^{iqx} \langle 0 | {\cal T}[j(x)j^{\dagger}(0)]| 0\rangle, \label{eqn1}
\end{equation} 
where $j(x)$ is the interpolating current, $q$ is the momentum of the state and ${\cal T}$ is the time ordering operator. Currents are suitably chosen operators made of quark and gluon fields that can create the studied hadron from vacuum. 

The key ingredient in the correlation function is the $j(x)$ operator. If the structure of the operator resembles the structure of the hadron, then obtained sum rules are expected to be more reliable. 

In molecular picture being a $J^{PC}=1^{++}$ state, $X(3872)$ can be represented as
\begin{equation}
\vert X(3872)\rangle=\frac{1}{\sqrt{2}}(|D\bar{D}^*\rangle-|\bar{D}D^*\rangle).
\end{equation}
In terms of quark, one can decompose this state like
\begin{equation}
\vert X(3872)\rangle=\frac{1}{\sqrt{2}} (\vert ((c\bar{q})(0)(\bar{c}q)(1))(1)\rangle - \vert ((\bar{cq})(0)(c\bar{q})(1))(1)\rangle). 
\end{equation}
In this representation $(c\bar{q})(0)$ means that total spins of $c$ and  $\bar{q}$ quarks is equal to $(0)$. It is also possible to represent this state in terms of another basis in Hilbert space \cite{41}:
\begin{equation}
\vert X(3872)\rangle=\vert ((c\bar{c})(1)(q\bar{q})(1))(1)\rangle
= \vert (S_{c\bar{c}}=1;S_{q\bar{q}}=1) J=1\rangle.
\end{equation}
It can be seen from above equation that charm and light quarks have total spin 1, respectively. These particles came together having total spin 1. Two spin-1 systems can result total spin 0, 1 and 2. Spin-0 and spin-2 are denoted as $X(3872)$ partners.

In this paper, the current 
\begin{equation}
j_{\mu\nu}=\bar{Q}^a \gamma_\mu Q^b \bar{q}^b \gamma_\nu q^a \label{current}
\end{equation}
is used. This current was proposed in \cite{41} to study $X(3872)$ and its partners. For $X(3872)$, $Q=c$ and $q=u$ or $q=d$, and $a$ and $b$ are color indices. As is customary in the QCD sum rules and lattice literature, annihilation diagrams are ignored in this work. Also, the masses of the $u$ and $d$ quarks are taken to be zero. The color combination is chosen such that the current can create  colorless $D$ and $D^*$ states. This current has even charge parity, $C=+$. An advantage of this current is that, $d=3$ term in the OPE, i.e. the quark condensate term, does not contribute to the sum rules.

Using this current, the correlation function can be written as
\begin{eqnarray}
\Pi_{\alpha \beta \gamma \delta}&=& i \int d^4x e^{iqx} \langle 0 | T[j_{\alpha\beta}(x)j_{\gamma\delta} ^\dagger(0)]| 0\rangle \nonumber \\
&=& i \int d^4x e^{iqx}  \langle 0 | T \bar{Q}^a(x) \gamma_\alpha Q^b(x) \bar{q}^b(x) \gamma_\beta q^a(x)
 \times  \bar{Q}^c(0) \gamma_\gamma Q^d(0) \bar{q}^d(0) \gamma_\delta q^c(0)  \vert 0 \rangle.
\end{eqnarray}

Following \cite{41}, three projections operators are defined as
\begin{equation}
{\cal P}^2_{\mu \nu \bar{\mu} \bar{\nu}}= \frac{1}{2}(g_{\mu \bar{\mu}} g_{\nu \bar{\nu}}+ g_{\mu \bar{\nu}} g_{\nu \bar{\mu}}-\frac{1}{2}g_{\mu \nu} g_{\bar{\mu} \bar{\nu}}),
\end{equation}
\begin{equation}
{\cal P}^1_{\mu \nu \bar{\mu} \bar{\nu}}= \frac{1}{2}(g_{\mu \bar{\mu}} g_{\nu \bar{\nu}}- g_{\mu \bar{\nu}} g_{\nu \bar{\mu}}),
\end{equation}
\begin{equation}
{\cal P}^0_{\mu \nu \bar{\mu} \bar{\nu}}= \frac{1}{4}(g_{\mu \nu} g_{\bar{\mu} \bar{\nu}}).
\end{equation}

Using these operators, interpolating currents can be written as the sum of three irreducible representations of the Lorentz group as
\begin{equation}
j_{\mu \nu}= j_{\mu \nu}^{2+} + j_{\mu \nu}^{1+} + j_{\mu \nu}^{0+},
\end{equation}
where 
\begin{equation}
j_{\mu \nu}^{2+}= {\cal P}_{2 \mu \nu}^{\alpha \beta} j_{\alpha \beta}
=\frac{1}{2}(j_{\mu \nu}+j_{\nu \mu}-\frac{1}{2}g_{\mu \nu}j_\alpha^\alpha),
\end{equation}
\begin{equation}
j_{\mu \nu}^{1+}={\cal P}_{1 \mu \nu}^{\alpha \beta} j_{\alpha \beta}
=\frac{1}{2}(j_{\mu \nu} j_{\nu \mu}),
\end{equation}
\begin{equation}
j_{\mu \nu}^{0+}= {\cal P}_{0 \mu \nu}^{\alpha \beta} j_{\alpha \beta}= \frac{1}{4}g_{\mu \nu}j_\delta^\delta.
\end{equation}

In above equations, the superscript denotes the $J^C$ quantum numbers of the particle of largest spin that can be created by the corresponding operator. The $J^{PC}$ quantum numbers of the particles that can be created by these operators are as follows: $j_{\mu \nu}^{2+}$ can create $J^{PC}=0^{++}$, $J^{PC}=1^{++}$ and $J^{PC}=2^{++}$,  $j_{\mu \nu}^{1+}$ can create $J^{PC}=1^{++}$ and $J^{PC}=1^{-+}$, $j_{\mu \nu}^{0+}$ can create $J^{PC}=0^{++}$ from the vacuum. 

The phenomenological side of the correlation function obtained from $j^{2+}$ can be written as
\begin{eqnarray}
\Pi_{\mu \nu\alpha\beta}^{(2)} &=& i \int d^4x e^{i q x} \langle 0 \vert {\cal T} j^{2+}_{\mu \nu}(x) {j^{2+}_{\alpha \beta}}^\dagger (0) \vert 0 \rangle \nonumber \\
&=& \sum_h \frac{\langle 0 \vert j^{2+}_{\mu \nu} \vert h(q) \rangle \langle h(q) \vert j^{2+}_{\alpha \beta} \vert 0 \rangle}{q^2 - m_h^2} \nonumber \\
&=&\frac{(\lambda_2^{2^{++}})^2}{q^2 - m_{2^{++}}^2} \sum_s \epsilon_{\mu\nu} \epsilon^*_{\alpha \beta}\nonumber \\
&+&\frac{(\lambda_2^{1^{++}})^2}{q^2 - m_{1^{++}}^2} \sum_s \left(\epsilon_\mu q_\nu + q_\mu \epsilon_\mu \right) \left( \epsilon^*_\alpha q_\beta +q_\alpha \epsilon^*_\beta \right) \nonumber \\
&+&\frac{(\lambda_2^{0^{++}})^2}{q^2 - m_{0^{++}}^2} \left( q_\mu q_\nu - \frac14 g_{\mu \nu} \right) \left(q_\alpha q_\beta - \frac14 g_{\alpha \beta} \right)
\end{eqnarray}
where $m_{J^{PC}}$ denotes the mass of the meson whose quantum numbers are $J^{PC}$, and summations are over the spins of the corresponding meson.

The constants $\lambda_2^{J^{PC}}$ are defined through the matrix elements
\begin{equation}
\langle 2^{++} \vert j^{2++}_{\mu\nu} \vert 0 \rangle = \lambda_2^{2++} \epsilon_{\mu \nu},
\end{equation}
\begin{equation}
\langle 1^{++} \vert j^{2++}_{\mu\nu} \vert 0 \rangle = \lambda_2^{1^{++}} \left(\epsilon_\mu q_\nu + \epsilon_\nu q_\mu\right),
\end{equation}
\begin{equation}
\langle 0^{++} \vert j^{2++}_{\mu\nu} \vert 0 \rangle = \lambda_2^{0^{++}} \left( \frac {q_\mu q_\nu}{q^2} - \frac{1}{4} g_{\mu\nu} \right),
\end{equation}
where  $\epsilon_{\mu \nu}$ and $\epsilon_\mu$ are polarization tensors for spin-2 and spin-1 respectively and $q$ is the momentum of the hadron. The polarization tensors satisfy $q^\mu \epsilon_\mu=0$, $\epsilon_\mu \epsilon^{\mu \star}=-1$, $q^\mu \epsilon_{\mu \nu}=0 $, $\epsilon_{\mu \nu}= \epsilon_{\nu \mu}$, $\epsilon_{\mu \nu} g^{\mu \nu}=0$ and $\epsilon_{\mu \nu} \epsilon^{\mu \nu \star}=1$. Polarization sum can be done via
\begin{eqnarray}
&& \sum_{s} \epsilon_{\mu \nu} \epsilon^*_{\alpha \beta}
\nonumber \\
&=&  \frac12 \left[
 \left(g_{\mu \alpha} - \frac{q_\mu q_{\alpha}}{q^2} \right)
 \left(g_{\nu \beta} - \frac{q_\nu q_\beta}{q^2} \right) 
 \right. \nonumber \\ && \left. +
 \left(g_{\nu \alpha} - \frac{q_\nu q_{\alpha}}{q^2} \right) 
 \left(g_{\mu \beta} - \frac{q_\mu q_\beta}{q^2} \right)
 \right. \nonumber \\
 &&\left. 
 - \frac 23 \left(g_{\mu \nu} - \frac{q_\mu q_\nu}{q^2} \right) \left( g_{\alpha \beta} - \frac{q_\alpha q_\beta}{q^2} \right)
  \right]
\end{eqnarray}
for spin-2 mesons, and
\begin{equation}
\sum_s \epsilon_\mu \epsilon^*_\nu = - \left( g_{\mu \nu} - \frac{q_\mu q_\nu}{q^2} \right) \equiv - g_{\mu \nu}^\perp
\end{equation}

for spin-1 mesons. For simplicity, defining the spin sums as

\begin{eqnarray}
\kappa^{22}_{\mu \nu \alpha \beta} &=&\sum_{s} \epsilon_{\mu \nu} \epsilon^*_{\alpha \beta}
\nonumber \\
&=&  \frac12 \left[ g_{\mu \alpha}^\perp g_{\nu \beta}^\perp + g_{\nu \alpha}^\perp g_{\mu \beta}^\perp - \frac23 g_{\mu \nu}^\perp g_{\alpha \beta}^\perp \right],\\ 
 \kappa^{21}_{\mu \nu \alpha \beta} &=& \sum_s \left(\epsilon_\mu q_\nu + q_\mu \epsilon_\nu \right) \left( \epsilon^*_\alpha q_\beta + q_\alpha \epsilon^*_\beta \right) \nonumber \\
 &=&  - \left( g_{\mu \alpha}^\perp q_\nu q_\beta + g_{\mu \beta}^\perp q_\nu q_\alpha + g_{\nu \alpha}^\perp q_\mu q_\beta +  g_{\nu \beta}^\perp q_\mu q_\alpha \right), \\
\kappa^{20}_{\mu \nu \alpha \beta} &=& \left( \frac{q_\mu q_\nu}{q^2} - \frac14 g_{\mu \nu} \right) \left(\frac{q_\alpha q_\beta}{q^2} - \frac14 g_{\alpha \beta} \right),
\end{eqnarray}

 the correlation function can be written as

 \begin{eqnarray}
\Pi_{\mu \nu \alpha \beta}^{(2)} = 
\frac{(\lambda_2^{2^{++}})^2}{q^2 - m_{2^{++}}^2} \kappa^{22}_{\mu \nu \alpha \beta}
+\frac{(\lambda_2^{1^{++}})^2}{q^2 - m_{1^{++}}^2} \kappa^{21}_{\mu \nu \alpha \beta} \\ \nonumber 
+\frac{(\lambda_2^{0^{++}})^2}{q^2 - m_{0^{++}}^2} \kappa^{20}_{\mu \nu \alpha \beta}.
\end{eqnarray}

Observing that
\begin{equation}
{\kappa^{2i}}_{\mu \nu \alpha \beta} {\kappa^{2j;}}^{\mu \nu \alpha \beta} = 0, ~~\mbox{if $i \neq j$}.
\end{equation}
the contribution of each $J^{PC}$ particle to the correlation function can be extracted as
\begin{eqnarray}
\frac{(\lambda_2^{2^{++}})^2}{q^2 - m_{2^{++}}^2} = \frac15\kappa^{22}_{\mu \nu \alpha \beta} \Pi^{(2)\mu \nu \alpha \beta} \label{eq:2130}\\
\frac{q^2 (\lambda_2^{1^{++}})^2}{q^2 - m_{1^{++}}^2} = \frac{1}{6q^2}\kappa^{21}_{\mu \nu \alpha \beta} \Pi^{(2)\mu \nu \alpha \beta} \label{eq:2131}\\
\frac{(\lambda_2^{0^{++}})^2}{q^2 - m_{0^{++}}^2} = \frac{16}{9}\kappa^{20}_{\mu \nu \alpha \beta} \Pi^{(2)\mu \nu \alpha \beta} \label{eq:2132}.
\end{eqnarray}

A similar analysis of the phenomenological side of the correlation function made of the $j^{1+}$ current can be carried out. By inserting a complete set of states between the interpolating currents, the correlation function can be written as

\begin{eqnarray}
\Pi_{\mu \nu \alpha \beta}^{(1)} &=& i \int d^4x e^{i q x} \langle 0 \vert {\cal T} j_{\mu \nu}^1(x) j_{\alpha \beta}^{1\dagger}(0) \vert 0 \rangle \nonumber \\
&=& \frac{\langle 0 \vert j_{\mu \nu}^1 \vert 1^{++} \rangle\langle 1^{++} \vert j_{\alpha \beta}^{1 \dagger} \vert 0 \rangle}{q^2 - m_{1^{++}}^2}
 \\ \nonumber && + \frac{\langle 0 \vert j_{\mu \nu}^1 \vert 1^{-+} \rangle\langle 1^{-+} \vert j_{\alpha \beta}^{1 \dagger} \vert 0 \rangle}{q^2 - m_{1^{-+}}^2}.
\label{eq:ek13}
\end{eqnarray}

Defining matrix elements as

\begin{eqnarray}
\langle 0 \vert j_{\mu \nu}^1 \vert 1^{++} \rangle &=& \lambda_1^{1^{++}} \left(q_\mu \epsilon_\nu - q_\nu \epsilon_\mu\right) \\
\langle 0 \vert j_{\mu \nu}^1 \vert 1^{-+} \rangle &=& \lambda_1^{1^{+-}} \epsilon_{\mu \nu \alpha \beta} q^\alpha {\epsilon'}^\beta
\end{eqnarray}
and using polarization sum, spin-1 correlation function becomes
\begin{eqnarray}
\Pi_{\mu \nu \alpha \beta}^{(1)} &=& 
\frac{ (\lambda_1^{1^{++}})^2}{q^2 - m_{1^{++}}^2} \sum_s \left(q_\mu \epsilon_\nu - q_\nu \epsilon_\mu\right)\left(q_\alpha \epsilon^*_\beta - q_\beta \epsilon^*_\alpha\right) \nonumber \\
&&+ \frac{ (\lambda_1^{1^{-+}})^2}{q^2 - m_{1^{-+}}^2} \sum_s \epsilon_{\mu \nu \bar \mu \bar \nu} q^{\bar \mu} {\epsilon'}^{\bar \nu} \epsilon_{\alpha \beta \bar  \alpha \bar \beta} q^{\bar \alpha} {\epsilon'}^{\bar \beta*}.
\label{eq:sr37}
\end{eqnarray}

Defining the  Lorentz structures
\begin{eqnarray}
\kappa^{1+}_{\mu\nu\alpha\beta} &=&\sum_s \left(q_\mu \epsilon_\nu - q_\nu \epsilon_\mu\right)\left(q_\alpha \epsilon^*_\beta - q_\beta \epsilon^*_\alpha\right)
\nonumber \\
&=&- \left( q_\mu q_\alpha g_{\nu\beta}^\perp - q_\mu q_\beta g_{\nu\alpha}^\perp - q_\nu q_\alpha g_{\mu\beta}^\perp + q_\nu q_\beta g_{\mu \alpha}^\perp \right),
\\
\kappa^{1-}_{\mu\nu\alpha\beta} &=&\sum_s \epsilon_{\mu \nu \bar \mu \bar \nu} q^{\bar \mu} {\epsilon'}^{\bar \nu} \epsilon_{\alpha \beta \bar  \alpha \bar \beta} q^{\bar \alpha} {\epsilon'}^{\bar \beta*} \nonumber \\
&=& -q^2 \left( g_{\mu \beta}^\perp g_{\nu \alpha}^\perp - g_{\mu \alpha}^\perp g_{\nu \beta}^\perp \right),
\end{eqnarray}
correlation function can be written in a  compact form as
\begin{eqnarray}
\Pi_{\mu\nu\alpha\beta}^{(1)} = \frac{ (\lambda_1^{1^{++}})^2}{q^2 - m_{1^{++}}^2}\kappa^{1+}_{\mu\nu\alpha\beta} 
+ \frac{ (\lambda_1^{1^{-+}})^2}{q^2 - m_{1^{-+}}^2} \kappa^{1-}_{\mu\nu\alpha\beta}.
\label{eq:sr40}
\end{eqnarray}

Using
\begin{equation}
\kappa^{1+}_{\mu\nu\alpha\beta} \kappa^{1-;\mu\nu\alpha\beta}=0,
\end{equation}
the contribution of the two particles can be isolated as

\begin{eqnarray}
\frac{ (\lambda_1^{1^{++}})^2}{p^2 - m_{1^{++}}^2} = \frac{1}{12}\kappa^{1+}_{\mu\nu\alpha\beta} \Pi_{\mu\nu\alpha\beta}^{(1)} \label{eq:2241}
\end{eqnarray} and
\begin{eqnarray}
\frac{  (\lambda_1^{1^{-+}})^2}{p^2 - m_{1^{-+}}^2}  = \frac{1}{12}\kappa^{1-}_{\mu\nu\alpha\beta} \Pi_{\mu\nu\alpha\beta}^{(1)} \label{eq:2242}.
\end{eqnarray}

Finally, to obtain the phenomenological representation of the correlation function composed of $j^0$, first note that $j^0$ can only create particles with quantum numbers $J^{PC}=0^{++}$. With the matrix element defined as
\begin{equation}
\langle 0 \vert j_{\mu \nu}^0 \vert 0^{++} \rangle = \lambda_0^{0^{++}} g_{\mu\nu},
\end{equation}

the correlation function can be written as follows

\begin{eqnarray}
\Pi_{\mu \nu \alpha \beta}^{(0)} &=& i \int d^4x e^{i q x} \langle 0 \vert {\cal T} j_{\mu \nu}^0(x) j_{\alpha \beta}^{0\dagger}(0) \vert 0 \rangle\nonumber \\ 
&=& \frac{(\lambda_0^{0^{++}})^2}{q^2 - m_{0^{++}}^2} g_{\mu\nu} g_{\alpha \beta},
\end{eqnarray}

which can be converted to

\begin{equation}
\frac{(\lambda_0^{0^{++}})^2}{q^2 - m_{0^{++}}^2} = \frac{1}{16} \Pi_{\mu\nu\alpha\beta}^{(0)} g^{\mu\nu}g^{\alpha\beta} \label{eq:2245}. 
\end{equation}

As can be seen from Eqs. \ref{eq:2130}-\ref{eq:2132}, \ref{eq:2241}, \ref{eq:2242}, and \ref{eq:2245}, the masses of the hadrons can all be obtained from equations of the form:
\begin{equation}
P(q^2) \frac{\lambda^2}{q^2 - m^2} = \Pi^{phen}(q^2) \label{eq:44},
\end{equation}
where $P(q^2)$ is a polynomial in $q^2$. Note that, the left hand side of Eq. \ref{eq:44} also contains contributions from higher states and the continuum, but only the contribution of the lowest state is explicitly written out.

 As is stated earlier, to match $\Pi^{phen}$ with $\Pi^{QCD}$, spectral representation of the correlation function is used:
\begin{equation}
\Pi(q^2) = \int_0^\infty ds \frac{\rho(s)}{s-q^2} + \mbox{polynomials is $q^2$},
\end{equation} 
where $\rho(s)$ is the spectral density. To get rid of the unknown polynomials, Borel transformation is carried out. After the Borel transformation, Eq. \ref{eq:44} can be written as:
\begin{equation}
P(m^2) \lambda^2 e^{- \frac{m^2}{M^2}} + \cdots = \int_0^\infty ds e^{-\frac{s}{M^2}}\rho^{QCD}(s), 
\end{equation}
where $M^2$ is the Borel parameter, and $\cdots$ represent the contributions of the higher states and continuum.

To subtract the contributions of the higher states and the continuum, quark hadron duality is used. The motivation of quark hadron duality is that higher states are to be found at higher energies where OPE can be used. In that region, for above a critical energy spectral function of the continuum and higher states can be represented as the spectral function of OPE. In quark hadron duality, it is assumed the $\rho^{phen}(s) = \rho^{QCD}(s)$ for $s > s_0$, where $s_0$ is called the continuum threshold. After using quark hadron duality, the sum rules can be obtained as
\begin{equation}
P(m^2) \lambda^2 e^{- \frac{m^2}{M^2}}  = \int_0^{s_0} ds e^{-\frac{s}{M^2}}\rho^{QCD}(s).
\end{equation}

The mass of the relevant meson can be obtained from the sum rules by taking the derivative of the logarithm of both sides with respect to $1/M^2$ as:
\begin{equation}
m^2 = \frac{\int_0^{s_0} ds e^{-\frac{s}{M^2}}s \rho^{QCD}(s) }{\int_0^{s_0} ds e^{-\frac{s}{M^2}}.\rho^{QCD}(s) }
\end{equation}

The analytical expressions for $\rho^{QCD}(s)$ are presented in the Appendix.

\section{\label{sec:level3}Numerical Analysis of Mass Spectrum}

The numerical values for QCD parameters used in this work are $m_c = 1.4 ~ GeV^{2}$, $m_b=4.7 ~  GeV^{2}$, $m_u =m_d = 0$, and $\frac{1}{4 \pi^2} \langle g_s^2 G^2 \rangle= 0.012 ~  GeV^{4}$. There are two additional  parameters in QCD Sum Rule calculations. These are the Borel parameter (or Borel mass) and continuum threshold. Borel parameter $M^2$, is an auxiliary parameter so physical properties should not depend on it. Due to the approximation made, a residual dependence on $M^2$ exist. Hence, a range for the Borel parameter in which physical observations are independent of it should be found. The other parameter is continuum threshold, $s_0$. In general, this parameter is taken to be $s_0\simeq (m+0.5~GeV)^2$ where $m$ denotes the mass of the studied hadron.

In the present work, the results of the sum rules are studied for the two values of continuum threshold:  $s_0= 17~ GeV^2$ and $s_0= 19~ GeV^2$ for $X(3872)$ and its partners in the charm sector, and $s_0=100~GeV^2$ or $s_0=102~GeV^2$ for the bottom sector.

\begin{figure}[H]
\centering
\includegraphics[width=3.4in]{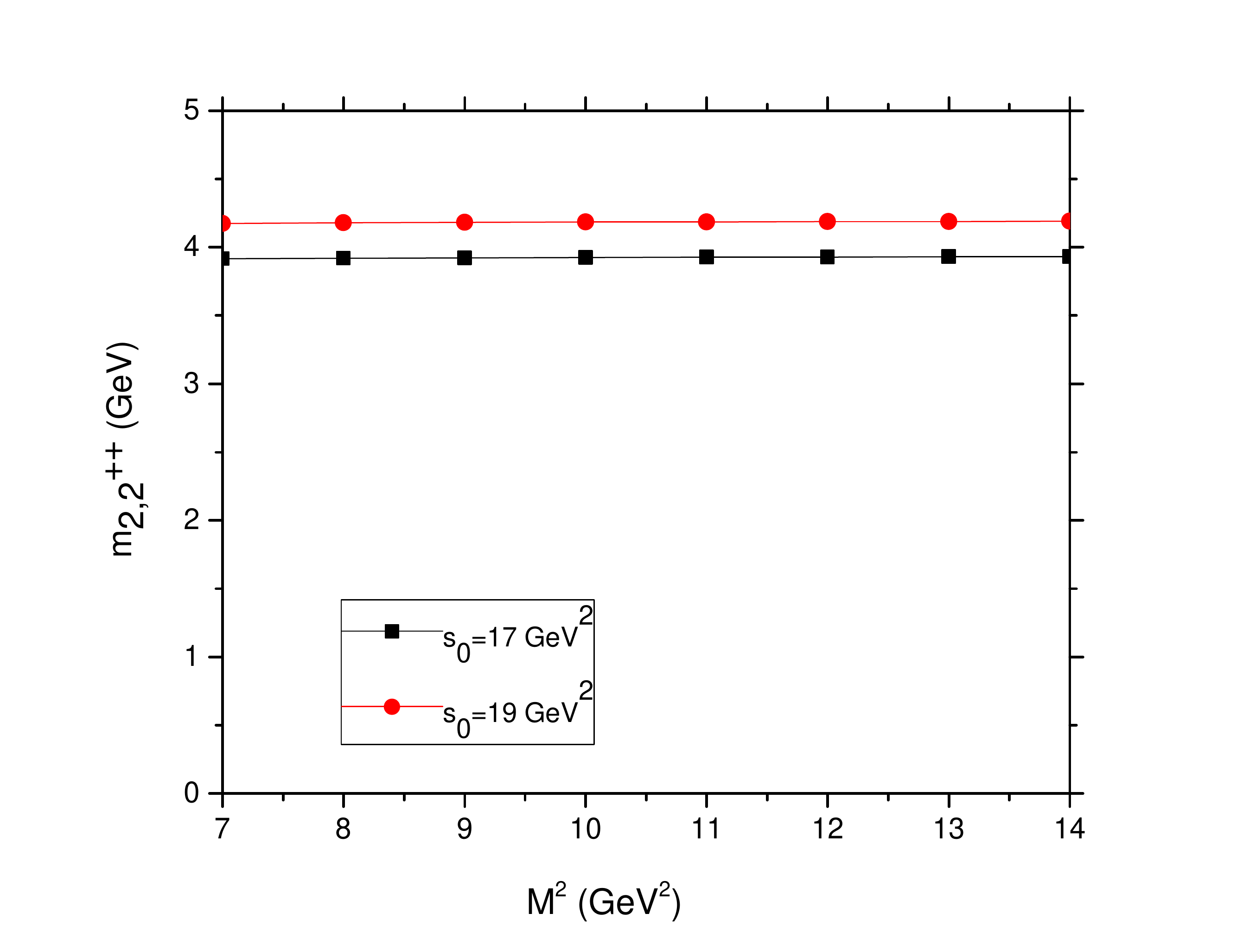}
\caption{\label{fig:Fig1} Borel parameter dependence of $J^{PC}=2^{++}$ meson mass from $\Pi_{\mu \nu \alpha \beta}^{(2)}$ for different $s_0$ values} 
\end{figure}

\begin{figure}[H]
\centering
\includegraphics[width=3.4in]{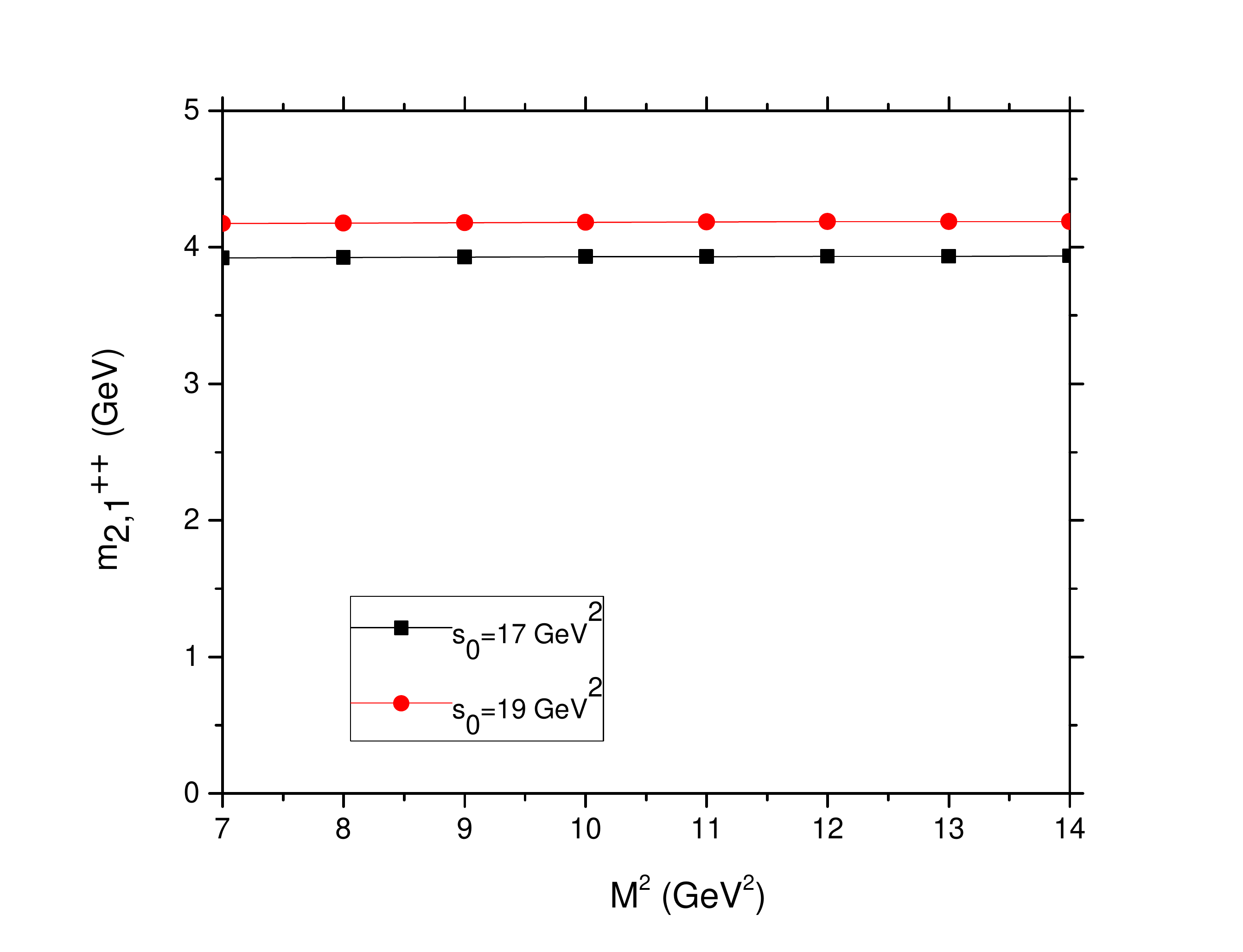}
\caption{\label{fig:Fig2} Borel parameter dependence of $J^{PC}=1^{++}$ meson mass from $\Pi_{\mu \nu \alpha \beta}^{(2)}$ for different $s_0$ values} 
\end{figure}

\begin{figure}[H]
\centering
\includegraphics[width=3.4in]{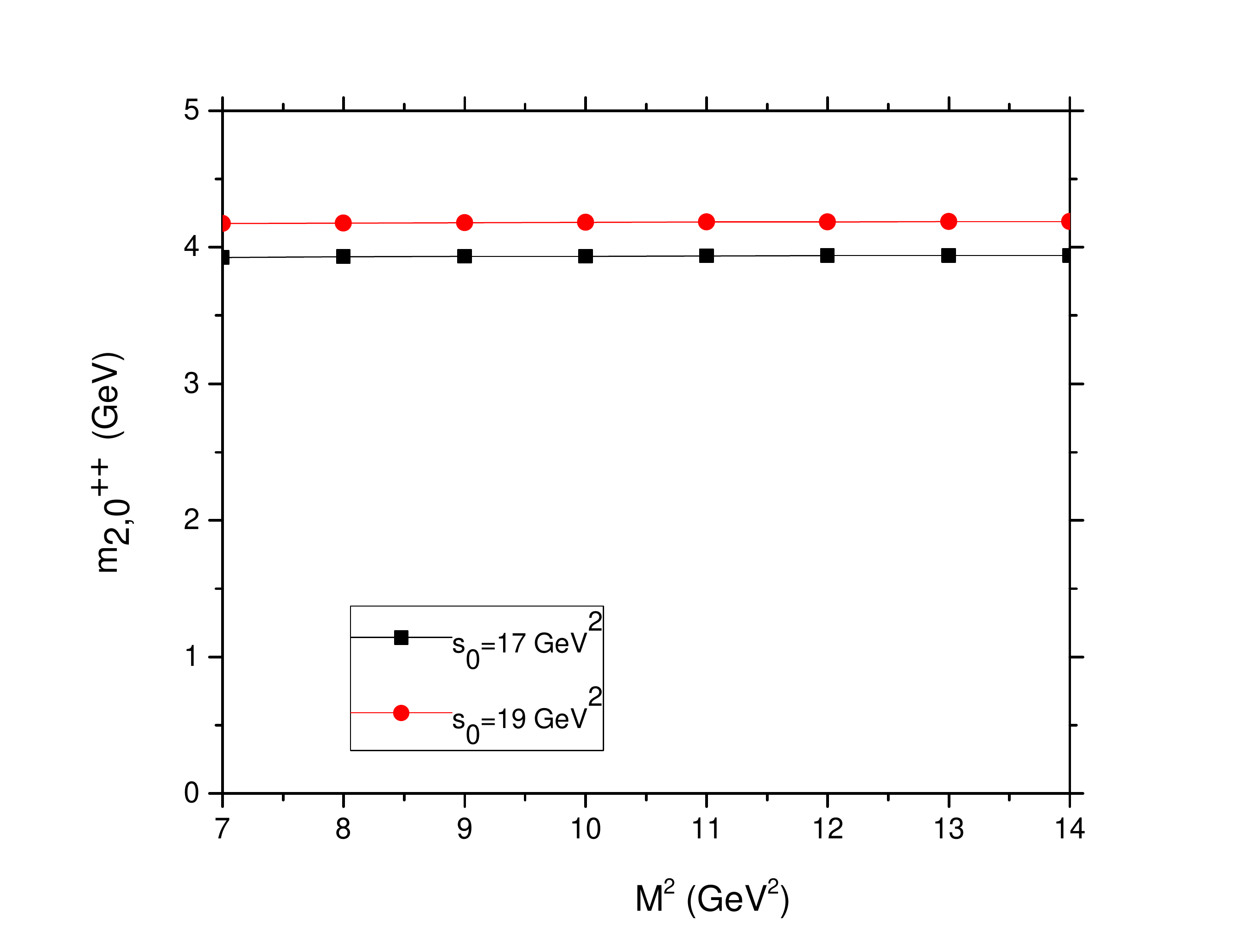}
\caption{\label{fig:Fig3} Borel parameter dependence of $J^{PC}=0^{++}$ meson mass from $\Pi_{\mu \nu \alpha \beta}^{(2)}$ for different $s_0$ values} 
\end{figure}

\begin{figure}[H]
\centering
\includegraphics[width=3.4in]{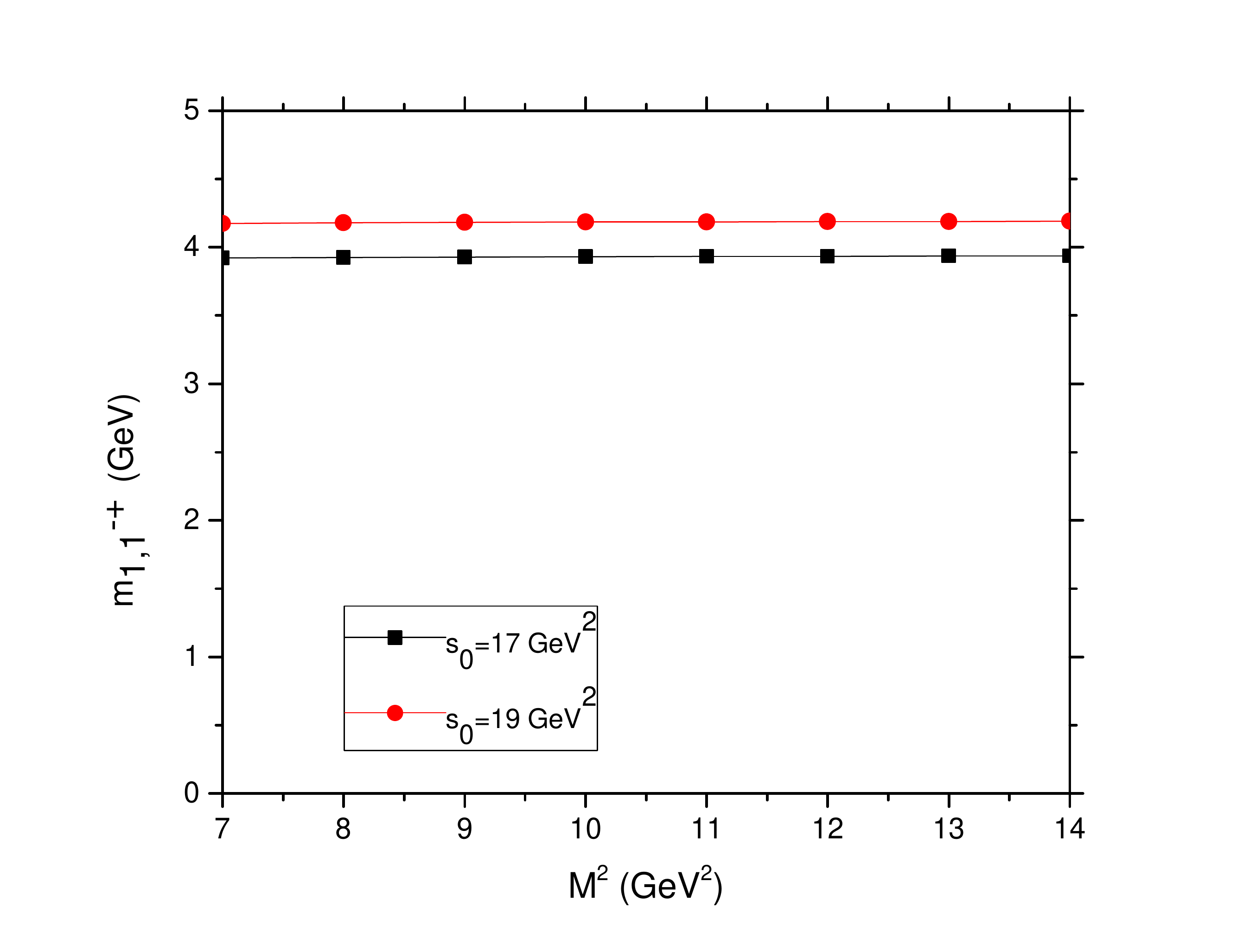}
\caption{\label{fig:Fig4} Borel parameter dependence of $J^{PC}=1^{-+}$ meson mass from $\Pi_{\mu \nu \alpha \beta}^{(1)}$  for different $s_0$ values} 
\end{figure}

\begin{figure}[H]
\centering
\includegraphics[width=3.4in]{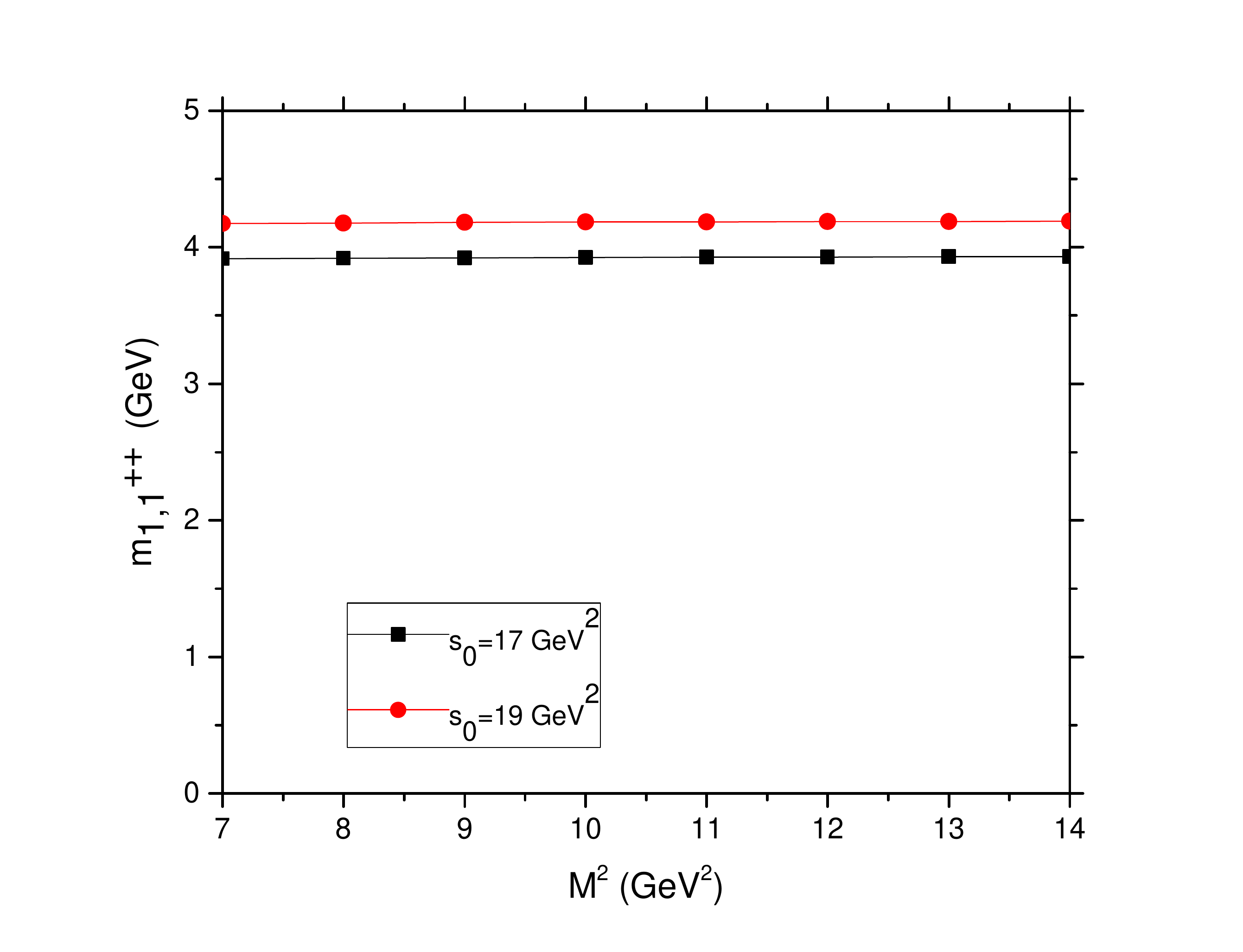}
\caption{\label{fig:Fig5} Borel parameter dependence of $J^{PC}=1^{++}$ meson mass from $\Pi_{\mu \nu \alpha \beta}^{(1)}$ for different $s_0$ values} 
\end{figure}

\begin{figure}[H]
\centering
\includegraphics[width=3.4in]{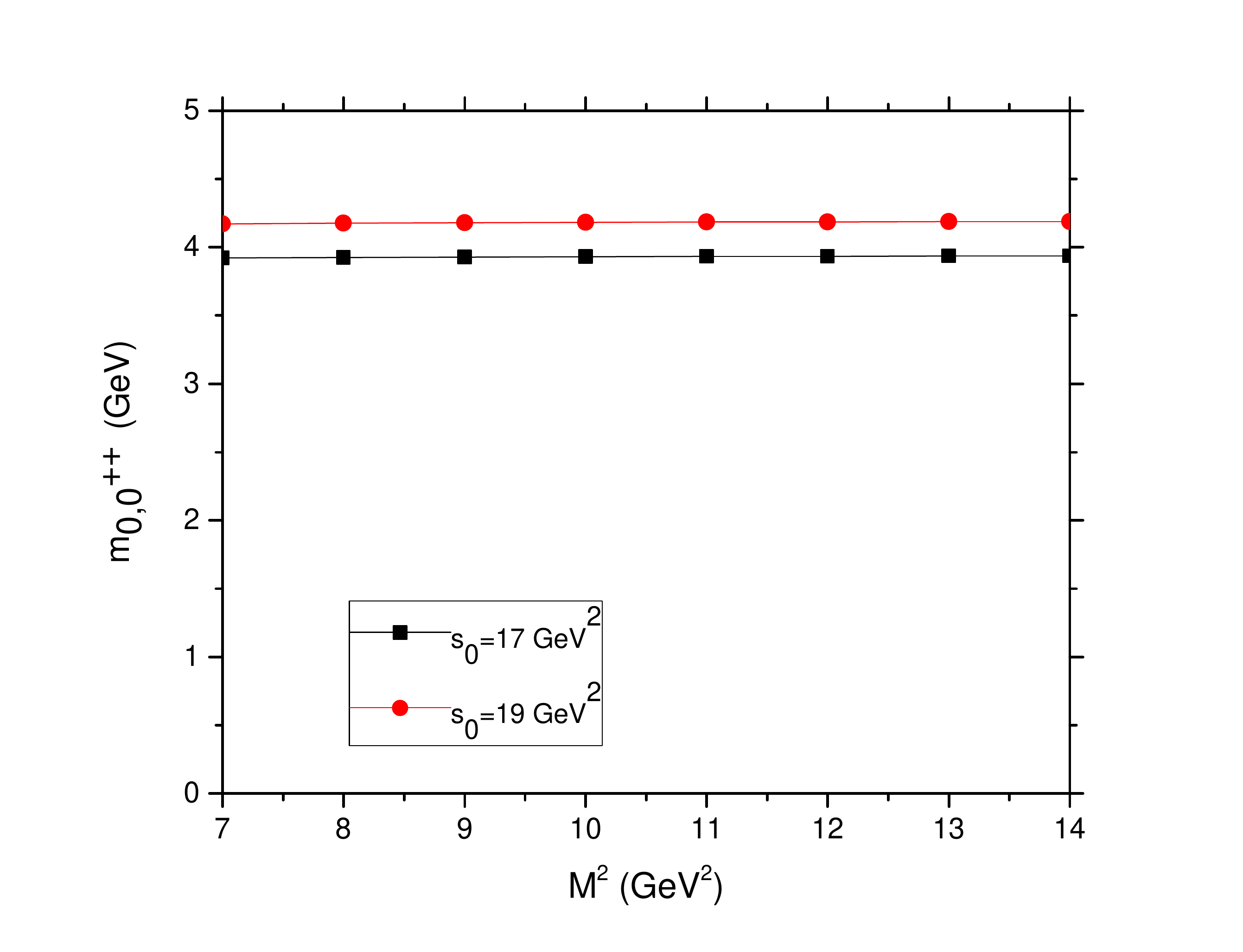}
\caption{\label{fig:Fig6}  Borel parameter dependence of  $J^{PC}=0^{++}$ meson mass from $\Pi_{\mu \nu \alpha \beta}^{(0)}$ for different $s_0$ values } 
\end{figure}

\begin{figure}[H]
\centering
\includegraphics[width=3.4in]{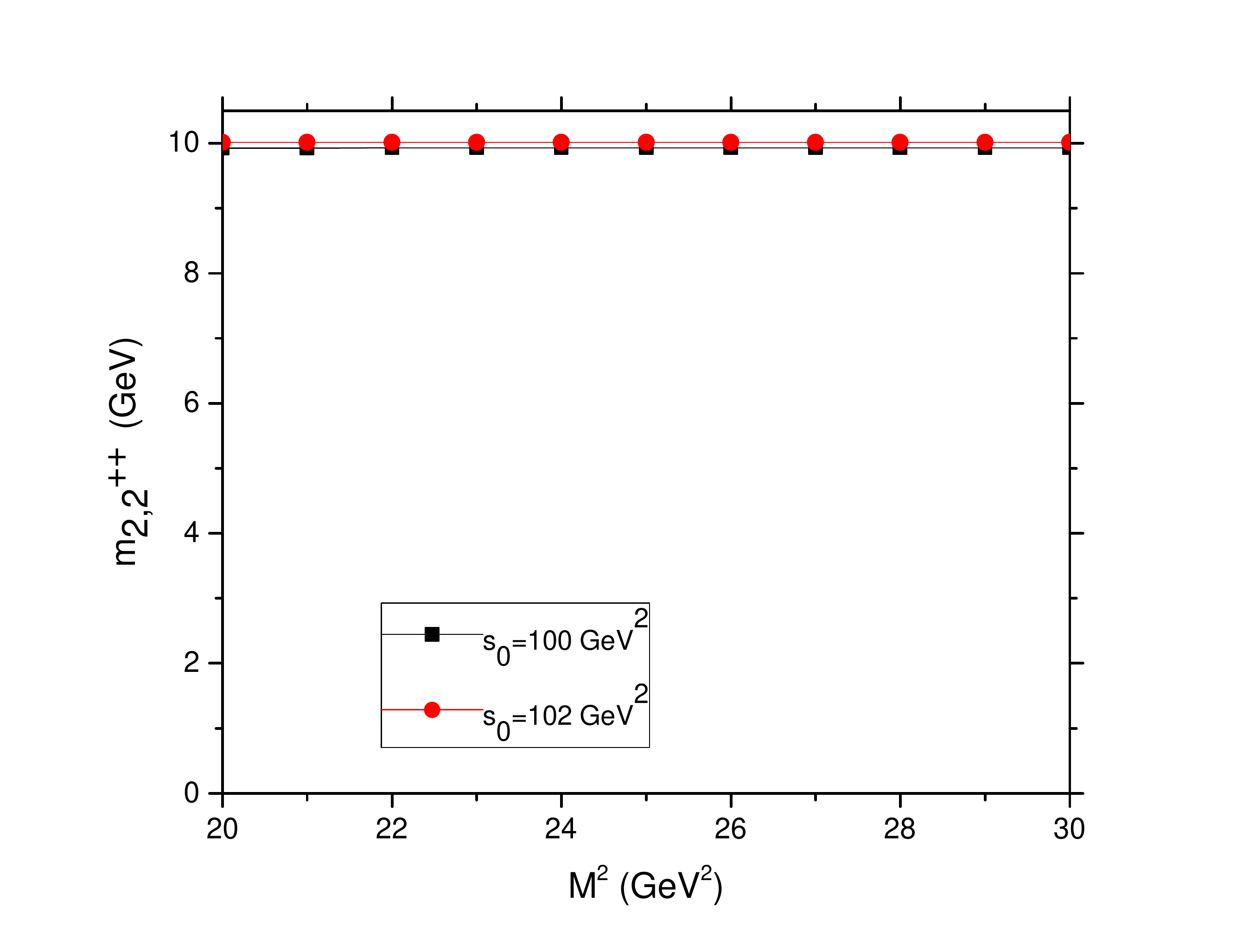}
\caption{\label{fig:Fig7} Borel parameter dependence of $J^{PC}=2^{++}$  meson mass from $\Pi_{\mu \nu \alpha \beta}^{(2)}$  for different $s_0$ values} 
\end{figure}

\begin{figure}[H]
\centering
\includegraphics[width=3.4in]{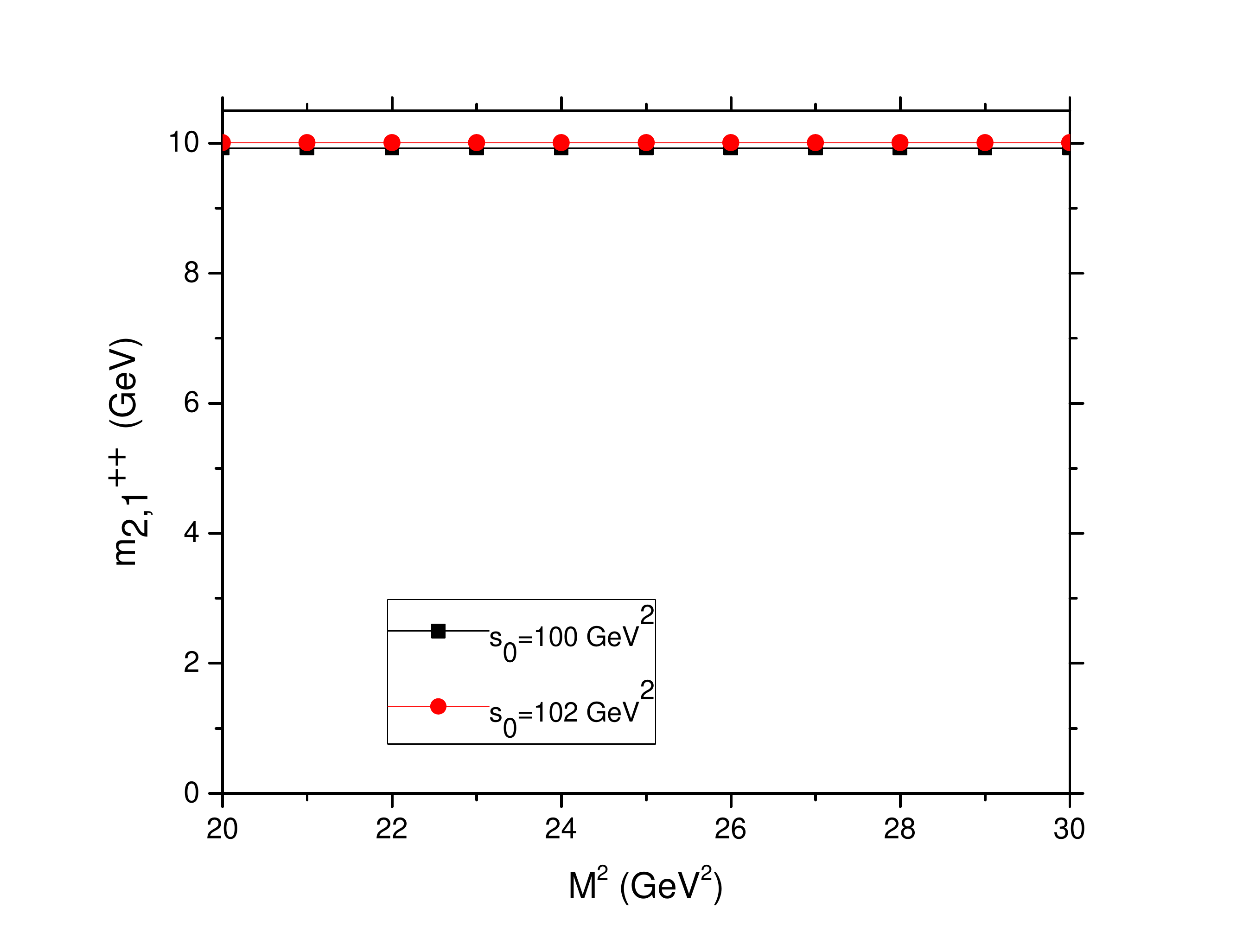}
\caption{\label{fig:Fig8} Borel parameter dependence of $J^{PC}=1^{++}$ meson mass from $\Pi_{\mu \nu \alpha \beta}^{(2)}$  for different $s_0$ values } 
\end{figure}

\begin{figure}[H]
\centering
\includegraphics[width=3.4in]{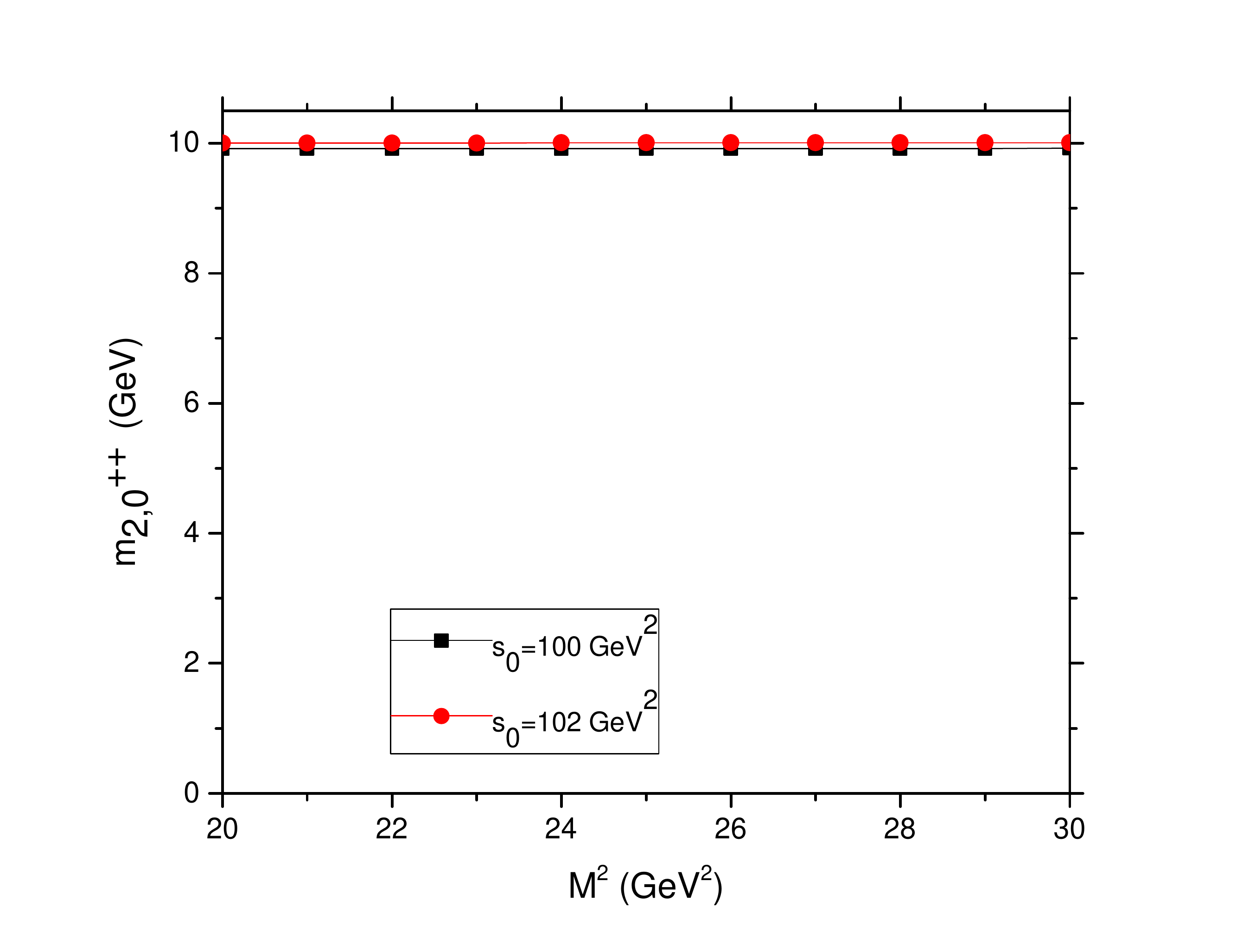}
\caption{\label{fig:Fig9} Borel parameter dependence of $J^{PC}=0^{++}$  meson mass from $\Pi_{\mu \nu \alpha \beta}^{(2)}$  for different $s_0$ values} 
\end{figure}

\begin{figure}[H]
\centering
\includegraphics[width=3.4in]{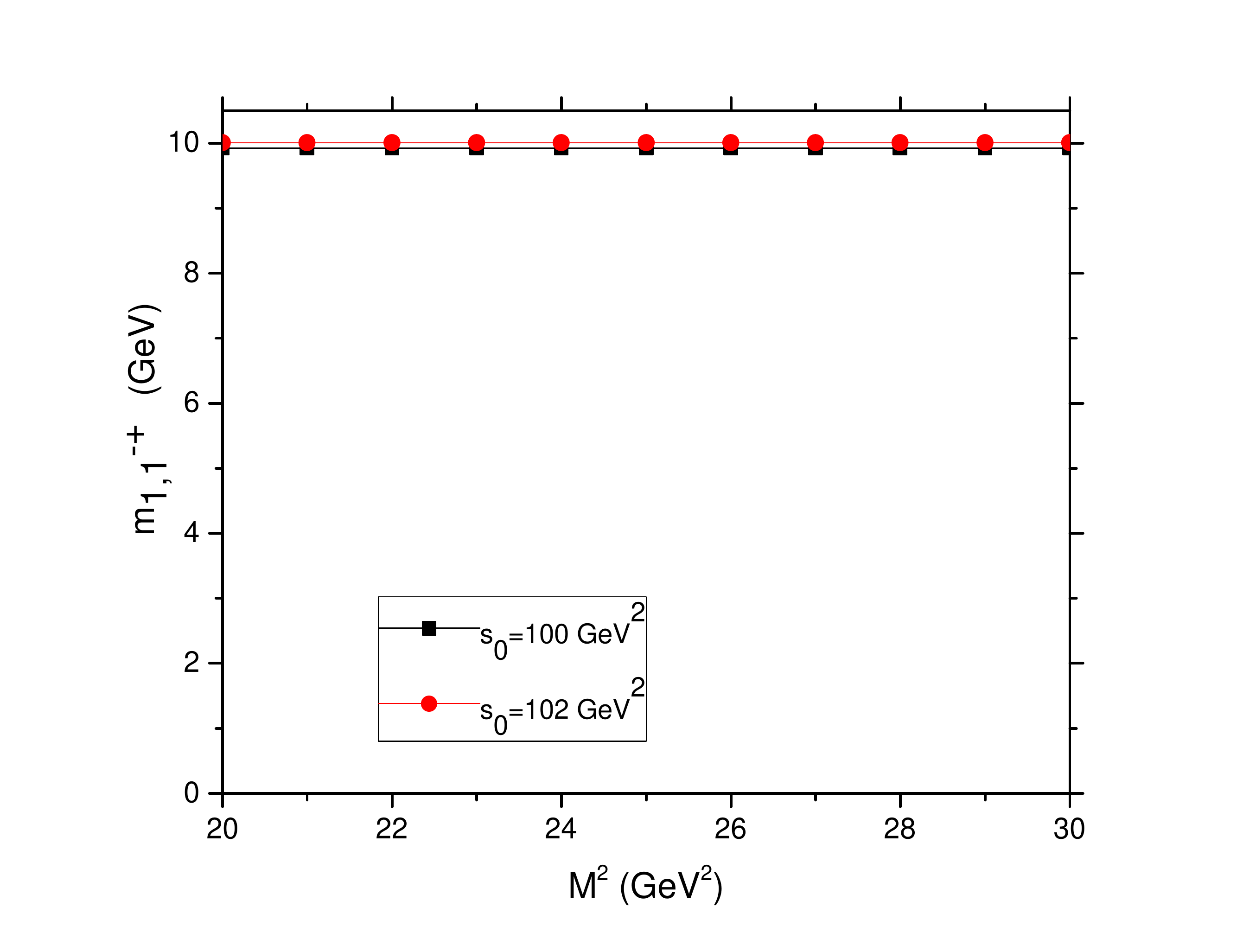}
\caption{\label{fig:Fig10} Borel parameter dependence of $J^{PC}=1^{-+}$ meson mass from $\Pi_{\mu \nu \alpha \beta}^{(1)}$ for different $s_0$ values} 
\end{figure}

\begin{figure}[H]
\centering
\includegraphics[width=3.4in]{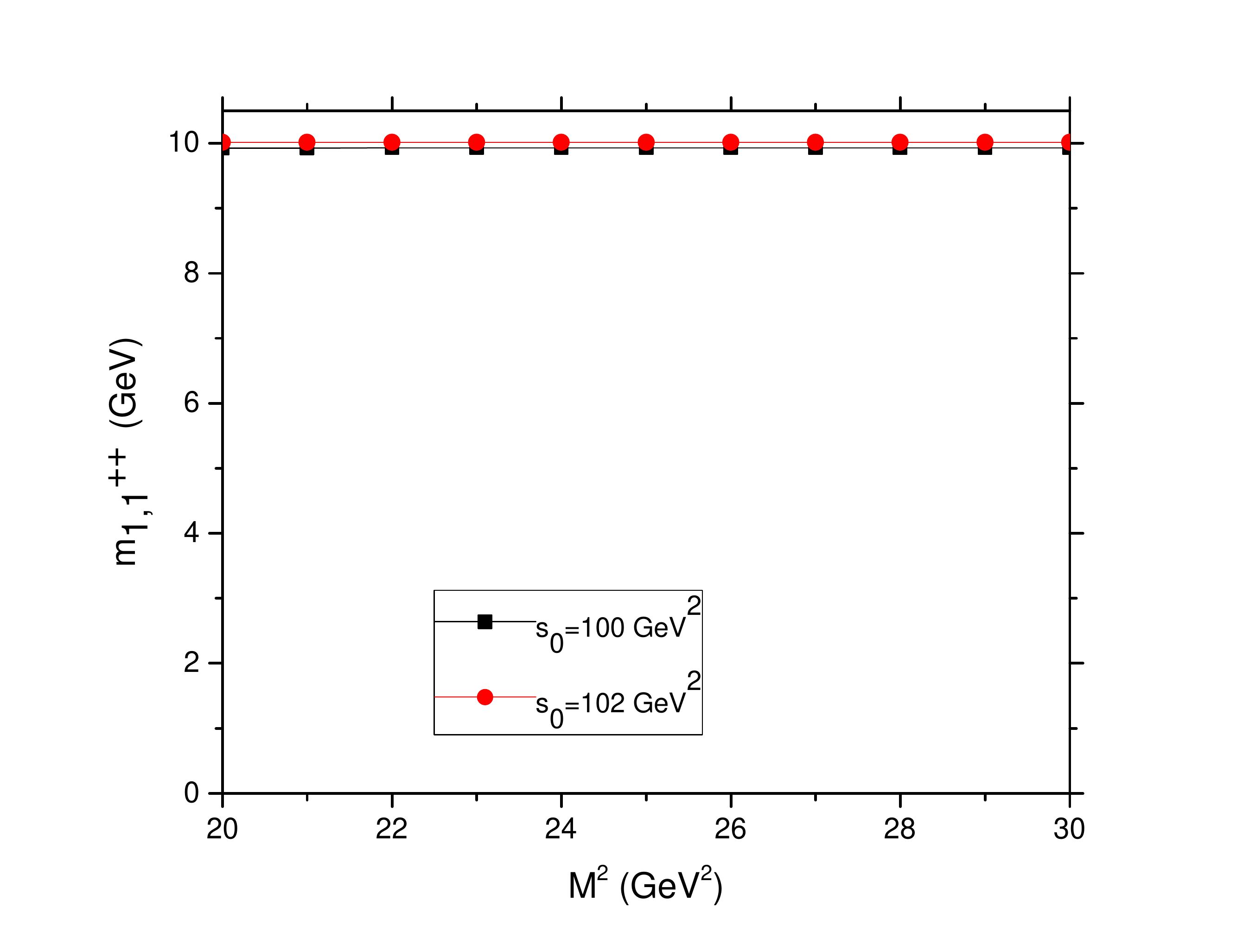}
\caption{\label{fig:Fig11} Borel parameter dependence of $J^{PC}=1^{++}$ meson mass from $\Pi_{\mu \nu \alpha \beta}^{(1)}$ for different $s_0$ values} 
\end{figure}

\begin{figure}[H]
\centering
\includegraphics[width=3.4in]{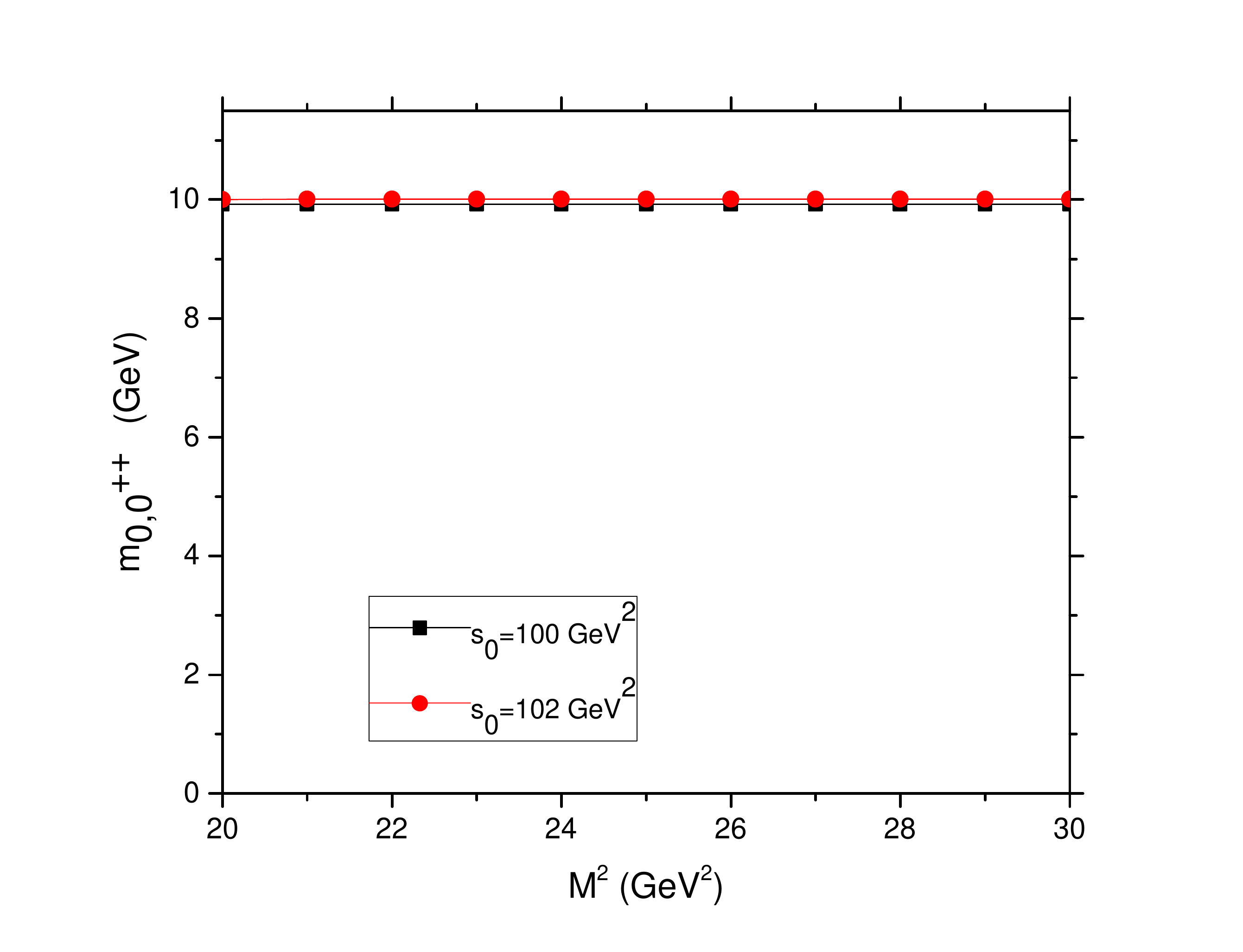}
\caption{\label{fig:Fig12} Borel parameter dependence of $J^{PC}=0^{++}$ meson mass from $\Pi_{\mu \nu \alpha \beta}^{(0)}$ for different $s_0$ values } 
\end{figure}

In the charm sector, the dependencies of the masses on the Borel parameter for the two values of the continuum threshold are shown in Figures \ref{fig:Fig1}, \ref{fig:Fig2}, \ref{fig:Fig3}, \ref{fig:Fig4}, \ref{fig:Fig5}, \ref{fig:Fig6}. Figures \ref{fig:Fig7}, \ref{fig:Fig8}, \ref{fig:Fig9}, \ref{fig:Fig10}, \ref{fig:Fig11} and \ref{fig:Fig12} are the same as Figures 1-6, but for the bottom sector. As can be observed from all the figures, the residual dependence on the Borel parameter is negligible for the chosen continuum thresholds in the chosen Borel range, which is an indication in favour of the chosen ranges. 

In Tables \ref{tab:table1} and \ref{tab:table2}  we present our predictions for the masses of the particles with $J^{PC}$ quantum numbers $0^{++}$, $1^{++}$, $1^{-+}$ and $2^{++}$ in the charm and bottom sector. The error bars in the table are mainly due to the variations of the prediction with the continuum threshold. Note that that $1^{-+}$ particle present in the tables is not the partner of $X(3872)$. $X(3872)$ corresponds to the $1^{++}$ state, and the predicted mass is higher than the experimental value. Note that in this section, all the obtained masses are almost degenerate with each other. 

\begin{table}[ht]
\caption{\label{tab:table1}Mass spectrum of $X(3872)$ partners}
\begin{ruledtabular}
\begin{tabular}{ccc}
Phenomenological side & $J^{PC}$ & $M_X $  \\
\hline
$\Pi_{\mu \nu \alpha \beta}^{(0)}$ & $0^{++}$ & $4055 \pm 126 ~ MeV $  \\
\hline
$\Pi_{\mu \nu \alpha \beta}^{(1)}$ & $1^{-+}$ & $4056 \pm 126 ~ MeV $ \\
& $1^{++}$ & $4053 \pm 129 ~ MeV $  \\
\hline
$\Pi_{\mu \nu \alpha \beta}^{(2)}$ & $0^{++}$ & $4058 \pm 124 ~ MeV $  \\
 & $1^{++}$ & $4055 \pm 126 ~ MeV $ \\
 & $2^{++}$ & $4053 \pm 129 ~ MeV $\\

\end{tabular}
\end{ruledtabular}

\end{table}

\begin{table}[ht]
\caption{\label{tab:table2}Mass spectrum of $\bar{b}b$ partners}
\begin{ruledtabular}
\begin{tabular}{ccc}
Phenomenological side  & $J^{PC}$ & $M_X $  \\
\hline
$\Pi_{\mu \nu \alpha \beta}^{(0)}$ & $0^{++}$ & $9922 \pm 41 ~ MeV $  \\
\hline
$\Pi_{\mu \nu \alpha \beta}^{(1)}$ & $1^{-+}$ & $9927 \pm 42 ~ MeV $  \\
& $1^{++}$ & $9923 \pm 42 ~ MeV $ \\
\hline
$\Pi_{\mu \nu \alpha \beta}^{(2)}$ & $0^{++}$ & $9920 \pm 44 ~ MeV $  \\
& $1^{++}$ & $9923 \pm 44 ~ MeV $   \\
 & $2^{++}$ & $9927 \pm 44 ~ MeV $  \\

\end{tabular}
\end{ruledtabular}

\end{table}

\section{Conclusion \label{sec:level4}}
In this work, we obtained mass spectrum of heavy quark partners of $X(3872)$ and $b\bar{b}$ in QCDSR framework by modelling it as molecular state. The current we studied has an advantage to study  $X(3872)$ partners comparing to for example \cite{48,49,50}. The reason for that is there can be light hadrons that can have the same quantum numbers with exotic hadrons. In the case of $X(3872)$, $\chi_{c1}$(1P) has the mass $m=3510.66 \pm 0.007$ MeV with the quantum number $J^{PC}=1^{++}$. The pure $c\bar{c}$ operator in the current can cause this problem \cite{48}. The other reason is that the current we used have an advantage of studying the partners of the $X(3872)$ meson on an equal footing \cite{49,50}.

In computations we have employed QCD two-point sum rule method and take into account vacuum condensates only with dimension four. We have found central values for the mass of $X(3872)$ as $m_X=4055 \pm 127$ MeV and for the mass of $b\bar{b}$ as $m_{b\bar{b}}=9924 \pm 43$ MeV . In the case of $X(3872)$, the central values is higher roughly $50$ MeV than the observed mass and for bottomonium case, the central value agree well with available experimental data  and theoretical predictions within the errors. 

The masses for correlation functions can be found in Tables \ref{tab:table1} and \ref{tab:table2} for $X(3872)$ and its partners and $b\bar{b}$, respectively.  As can be seen from Fig. \ref{fig:Fig1}, \ref{fig:Fig2}, \ref{fig:Fig3}, \ref{fig:Fig4}, \ref{fig:Fig5}, \ref{fig:Fig6}, \ref{fig:Fig7}, \ref{fig:Fig8}, \ref{fig:Fig9}, \ref{fig:Fig10}, \ref{fig:Fig11} and \ref{fig:Fig12} the mass spectrum for the $X(3872)$ and $\bar{b}b$ and their partners don't depend on the Borel parameter. The degeneracies of $X(3872)$ and $\bar{b}b$  partners are compatible with the heavy quark spin symmetry prediction. Table \ref{tab:table3} compares our results for $X(3872)$ with other studies.

\begin{table}[H]
\caption{\label{tab:table3}$X(3872)$ partners. All results are in MeV.}
\begin{ruledtabular}
\begin{tabular}{cccccc}
$J^{PC}$ & $M_X $ (this work) & \cite{33} (SU(2) isoscalar) & \cite{33}(SU(2) isovector) &\cite{34} & \cite{34} (OPE Potential inc.)\\
\hline
 $0^{++}$ & $4055 \pm 126$ & $3712^{+11}_{-13}$ & $3733^{+1.1}_{-6.9}$ &  $3709^{+12}_{-14}$  & $3714$ \\
\hline
$1^{++}$ & $4053 \pm 129 $ & $3872$ (Input) & - &  $3872$ (Input)   & $3872$ (Input) \\
$1^{-+}$ & $4056 \pm 126$  & - &- & - & - \\
\hline
$0^{++}$ & $4058 \pm 124$ & $3907$ (Input) & - & $3907$ (Input) & $3907$ (Input) \\
$1^{++}$ & $4055 \pm 126$ & - & - & - & - \\
$2^{++}$ & $4053 \pm 129$ &  $4013^{?}_{-11}$ & - & $4012^{+4}_{-6}$ & $4015$ \\
\end{tabular}
\end{ruledtabular}
\end{table}
It can be seen from Table \ref{tab:table3} that, our results agree within the errors with reference studies. 

To identify a state whether it is a hadronic molecule or not, some further investigation is needed. In \cite{57}, the authors reviewed to identify hadronic molecules according to the Weinberg compositeness criterion, the pole counting approach, pole trajectories and generalization to resonances. All these expressions are equivalent. To mention Weinberg compositeness criterion in different view of point, one can write a meson wave function such as

\begin{equation}
\Psi_{Meson}=a\vert q\bar{q} \rangle + b \vert q\bar{q}g \rangle + c \vert qq\bar{q}\bar{q} \rangle + \cdots .
\end{equation}
In conventional quark picture coefficient $a$, in hybrid picture coefficient $b$, and in tetraquark and meson molecules the coefficient $c$ is dominant \cite{58}. Having a nonzero value of $c$ will determine whether a state is composite or molecular. This includes the study of line shapes for the relating state. 

It is clear that,  the computation of mass alone does not allow us to make a conclusion on the internal structure of $X(3872)$. Besides that when a state is first observed and existence of it still needs confirmation, QCDSR can be very useful for justifying this existence. It can provide evidence in favor or against the existence of state. In QCDSR formalism, one cannot deduce if a state have a tetraquark configuration or molecular configuration. Furthermore, the limitations in statements with QCDSR estimates come from uncertainties in the formalism \cite{18}.  It is stated in \cite{59} that, different proposed substructures (tetraquarks and molecules) lead to the same mass predictions within the accuracy of the method  indicating that the predictions of the $X$ meson mass is not sufficient for revealing its nature. 

Using the current in this study, it was proven in \cite{41} that states couple to Eqn. (\ref{current}) degenerate triplets with the quantum numbers $J^{PC}=2^{++}$, $J^{PC}=1^{++}$ and $J^{PC}=0^{++}$ which holds for any state that couples to the current independent of its internal structure in the heavy quark limit. One example for such triplet is $\chi_{b0}(9859)$, $\chi_{b1}(9892)$ and $\chi_{b2}(9912)$. The masses of this triplet differs from their average value by less than $30$ MeV. The other example is  $\chi_{c0}(3414)$, $\chi_{c1}(3510)$ and $\chi_{c2}(3556)$. Here, the mass difference is less then $80$ MeV. From these considerations one can arrive a result that for the $X(3872)$ should have spin-0 and spin-2 partners which have a mass difference at the order of $ \sim$ 100 MeV from the mass of $X(3872)$ and $\sim$ 50 MeV for bottomonium. Our results agree at the order of these differences. 

In a work of Matheus et al. \cite{60}, they studied nature of the meson $X(3872)$ by assuming to be an exotic four quark $(c \bar{c} q \bar{q})$ state with $J^{PC}=0^{++}$ quantum number. They found $m_X= 3925 \pm 127$ MeV and for the b-quark $m_{X_b}=10144 \pm 146$ MeV. Our center of mass values agree with these results within the errors. 

We obtained possible partners of $X(3872)$ and $b\bar{b}$ according to HPSS predictions by QCD Sum Rule method in molecular picture. Other methods can be studied for mass differences rather than direct mass calculations. Besides that $X(3872)$ production can be studied in different channels.

\begin{acknowledgements}
This research has been supported by TUBITAK (The Scientific and Technological Research Council of Turkey) under the grant no 114F234. H. Mutuk is so indebted to A. Ozpineci for maintaing a bursary under this project.
\end{acknowledgements}

\appendix*
\section{}

In this section, we give the spectral density expressions of correlation functions

\begin{eqnarray}
\rho_{2,2^{++}}&=&\int_{4m_c^2}^{s_0}ds e^{-s/M^2} \frac{1}{20s^2}[9(g_3+2g_4+g_6)+20g_5 \nonumber \\ &-&9s(g_3+2g_4+g_6)]\theta(s-s(x,y)) \\
\rho_{2,1^{++}}&=&-\int_{4m_c^2}^{s_0} ds e^{-s/M^2}\frac{1}{2s}[s (g_3+2g_4+g_6) \nonumber \\ &+&2(g_5+g_7)]\theta(s-s(x,y))\\
\rho_{2,0^{++}}&=&\int_{4m_c^2}^{s_0} ds e^{-s/M^2}\frac{1}{3}[48 s (g_1s+g_3+2g_4+g_6)\nonumber \\&+&64(g_5+g_7)]\theta(s-s(x,y))\\
\end{eqnarray}

\begin{eqnarray}
\rho_{1,1^{++}}&=&\int_{4m_c^2}^{s_0}ds e^{-s/M^2}\frac{1}{4}[s (-2(g_5-g_7)\nonumber \\  &+& (g_3 - 2g_4 + g_6) s)]\theta(s-s(x,y)) \\
\rho_{1,1^{+-}}&=&\int_{4m_c^2}^{s_0}ds e^{-s/M^2}\frac{1}{2}(g_5-g_7)s\theta(s-s(x,y))
\end{eqnarray}

\begin{eqnarray}
\rho_{0,0^{++}}&=&\int_{4m_c^2}^{s_0}ds e^{-s/M^2}\frac{1}{16}[4(4 g_8 + g_5 + g_7)\nonumber \\
 &+& s (8 g_2 + g_3 + 2g_4 + g_6 + g_1 s)]\theta(s-s(x,y))
\end{eqnarray}

where 

\begin{eqnarray}
g_1&=&\int_0^1 \int_0^{1-x} dx dy \frac{1}{256 \pi^6 t^8}3 x^3 y^3 z  \left(s x y z-m_c^2  w\right)^2 \\  
g_2&=&\int_0^1 \int_0^{1-x} dx dy \frac{1}{6144 \pi^6 t^8} x y [12 x y \left(m_c^2 p (x+y) - s x y z)\right)^3 \nonumber \\ &-& \pi ^2 \langle g^2G^2\rangle t^2 \left(m_c^2 p q (x+y)+3 r s x y\right)] \\      
g_3&=&\int_0^1 \int_0^{1-x} dx dy \frac{1}{1536 \pi^6 t^8 z}
 x^2 y^2 [9 x y(2 s z^2-m_c^2 p)(s x y z\nonumber \\ &-& m_c^2 p (x+y))^2 - \pi^2 \langle g^2G^2 \rangle m_c^2  t^3 u] \\
 g_4&=&\int_0^1 \int_0^{1-x} dx dy\frac{1}{6144 \pi^6 t^8} xy [12 x y (-s x yz + m_c^2 p (x + y))^3 \nonumber \\ &+& 
 \langle g^2G^2\rangle \pi^2 t^2 (3 r s x y + mc^2 p v (x + y))]\\
 g_5&=&\int_0^1 \int_0^{1-x} dx dy \frac{1}{12288 \pi^6 z t^8} \nonumber \\ & \times &
 [3 x y (s x y z - m_c^2 p (x + y))^4 \nonumber \\ &+& 
 \langle g^2G^2\rangle \pi^2 t^2  \nonumber \\ &\times &(-s x yz + 
    m_c^2 p (x + y)) (3 r s x y z + q mc^2 p (x + y))] \\ 
 g_6&=&\int_0^1 \int_0^{1-x} dx dy \frac{xyq}{1536 \pi^6 t^8} [\langle g^2G^2\rangle m_c^2 \pi^2 t^3 \nonumber \\ &+&(-3sxyz+m_c^2tq)^3] \\
 g_7&=&\int_0^1 \int_0^{1-x} dx dy \frac{1}{(12288 \pi^6 z^2 t^8} [-4\langle g^2G^2\rangle m_c^2 \pi^2 t^3\nonumber \\ &+& 3(-sxyz+m_c^2t)^3+ (utm_c^2+3sxyw) ] \\
 g_8&=&\int_0^1 \int_0^{1-x} dx dy \frac{1}{12288 \pi^6 z t^8}
 [-3xy(sxyz-m_c^2p(x+y))^4 \nonumber \\ &+& 
 \langle g^2G^2\rangle \pi^2 t^2 (-sxyz+m_c^2 p(x+y))]
\end{eqnarray}
and

\begin{eqnarray}
s(x,y)&=&m_Q^2 \left(\frac{x}{y}-\frac{2}{x+y-1}+\frac{y}{x}\right)\\
 p&=&(-1 + x) x + (-1 + x) y + y^2 \\
q&=& 12 (-1 + x)^2 x^2 + 
 x (24 + x (-45 + 17 x)) y \nonumber \\ &+& (12 + x (-45 + 13 x)) y^2 + (-24 + 
    17 x) y^3 + 12 y^4\\
r&=& 10 x^4 + x y (-3 + 4 y) (-2 + 5 y) + (-1 + y) y^2 (-3 + 10 y) \nonumber \\ &+& 
 x^3 (-13 + 20 y) + x^2 (3 + y (-23 + 28 y))\\
t&=&  x^2 + x (-1 + y) + (-1 + y) y\\
w&=&x^3 + 2 x (-1 + y) y + (-1 + y) y^2 + x^2 (-1 + 2 y)\\
u&=& -6 x^5 + 3 x^6 + x^2 (1 - 3 y) y^3 - 3 x (-1 + y) y^3 \nonumber \\ & +& 
 3 (-1 + y)^2 y^4 + x^3 y (3 + y - 8 y^2) - 3 x^4 (-1 + y + y^2)\\
v&=& 9 x^4 + 9 (-1 + y)^2 y^2 + x^2 (-1 + 3 y) (-9 + 8 y) \nonumber \\ & +& 
 x^3 (-18 + 17 y) + x (-1 + y) y (-18 + 17 y)\\
z&=& x+y-1
\end{eqnarray}

\end{document}